\documentclass[letterpaper,twocolumn,fleqn]{article} 

\usepackage{ist}
\usepackage{bbding} % checkmarks and xsolids
\usepackage{cite}                      % needed to automatically sort the references
\usepackage{url}
\usepackage{stfloats}

\usepackage{hyperref}
\hypersetup{
  colorlinks=true,
  linkcolor=blue,
  filecolor=blue,
  citecolor=black,
  urlcolor=cyan,
}

\graphicspath{{figures/}{pictures/}{images/}{./}} % where to search for the images

\title{FastPoints: A State-of-the-Art Point Cloud Renderer for Unity}

\author{Elias Neuman-Donihue, Michael Jarvis, Yuhao Zhu}

\date{} % date has an empty field.

\begin{document} 

\maketitle 

\thispagestyle{empty} % prevents the first page to be numbered

%%%%%%%%%%%%%%%%%%%%%%%%%%%%%%%%%%
% Abstract
%%%%%%%%%%%%%%%%%%%%%%%%%%%%%%%%%%

\begin{abstract}
In this paper, we introduce FastPoints, a state-of-the-art point cloud renderer for the Unity game development platform. Our program supports standard unprocessed point cloud formats with non-programmatic, drag-and-drop support, and creates an out-of-core data structure for large clouds without requiring an explicit preprocessing step; instead, the software renders a decimated point cloud immediately and constructs a shallow octree online, during which time the Unity editor remains fully interactive.
\end{abstract}

\section{Introduction}

Over the past decade, the point cloud data format has exploded in popularity. Acquisition devices, such as LiDAR and depth sensors, have become more accessible, and processing software~\cite{feng2020real} and hardware options have grown in number and quality~\cite{xu2019tigris, feng2020mesorasi, feng2022crescent}. Today, point clouds are being used in various fields, including medicine, archaeology~\cite{ye2022digital}, ecology,  robotics, and more. Users can gather their own point clouds with LiDAR cameras built into modern smartphones or ultra-maneuverable drones, download open-source conversion and analysis tools like LAStools~\cite{lastools} and pcl~\cite{pcl}, and view their clouds in free web-based viewers, such as Potree~\cite{schutz2016potree} and plas.io~\cite{plasio}, or in desktop programs like Unreal Engine or CloudCompare~\cite{cloudcompare}. 

This surge in popularity has introduced a host of aspiring new users of point clouds who have less technical experience than users of previous eras. These new users hail from various academic and industrial settings, and all have discovered new and exciting ways to apply point clouds to their respective disciplines. However, although point cloud acquisition and visualization technology has developed at a staggering pace, usability for these non-technical users has yet to receive the attention it demands.

One specific area where usability has lagged behind the state of the art is point cloud rendering. Once a user has acquired a point cloud through a laser scanner or RGB-D camera, they can choose from a wealth of rendering options for viewing it in a virtual scene, but all require certain compromises. 

Some provide seamless user experiences, allowing a user to easily and quickly view their point cloud files in a 3D environment, but lag behind modern advances in rendering performance and visual quality - specifically, many lack support for large point clouds, which are too big be loaded fully into memory and require an out-of-core data structure to load subsets of the point cloud dynamically based on camera position. 

Other renderers leverage recent developments in rendering technology and provide performant support for large clouds, but suffer from limited user interfaces and lack support for other graphical formats such as meshes of UI canvas elements. Additionally, these renderers often come with hefty preprocessing requirements, forcing users to download additional programs to convert their clouds to specific data structures and wait a considerable amount of time for their clouds to be processed before rendering.

In this paper, we present FastPoints, a new point cloud renderer for the Unity3D platform aimed at providing a modern, usable experience for non-technical point cloud users. FastPoints offers a number of benefits to user experience compared to other large point cloud viewers and Unity point cloud renderers:

\begin{itemize}
    \item Fully non-programmatic interface
    \item Integrable with standard mesh rendering pipelines
    \item Interactive during preprocessing
    \item Support for popular point cloud formats
\end{itemize}

A feature comparison of FastPoints and other state-of-the-art renderers can be seen in Table \ref{tab:features}. We open-source the implementation of FastPoints at \url{https://github.com/eliasnd/FastPoints} to gather community feedback and enable community-driven development.

\subsubsection{Fully Non-Programmatic Interface}

State-of-the-art renderers today, while powerful, generally require users to do some amount of programming to add their clouds to virtual scenes, either rewriting scripts to adjust scene parameters, downloading and running command line tools to process point clouds, or modifying variables to point the renderer to the user's clouds on disk. This can present a significant barrier to entry for non-technical users. FastPoints requires no programming knowledge, and all functionality is exposed via a graphical interface that integrates with the Unity editor.

\begin{figure}
    \centering
    \includegraphics[width=\columnwidth]{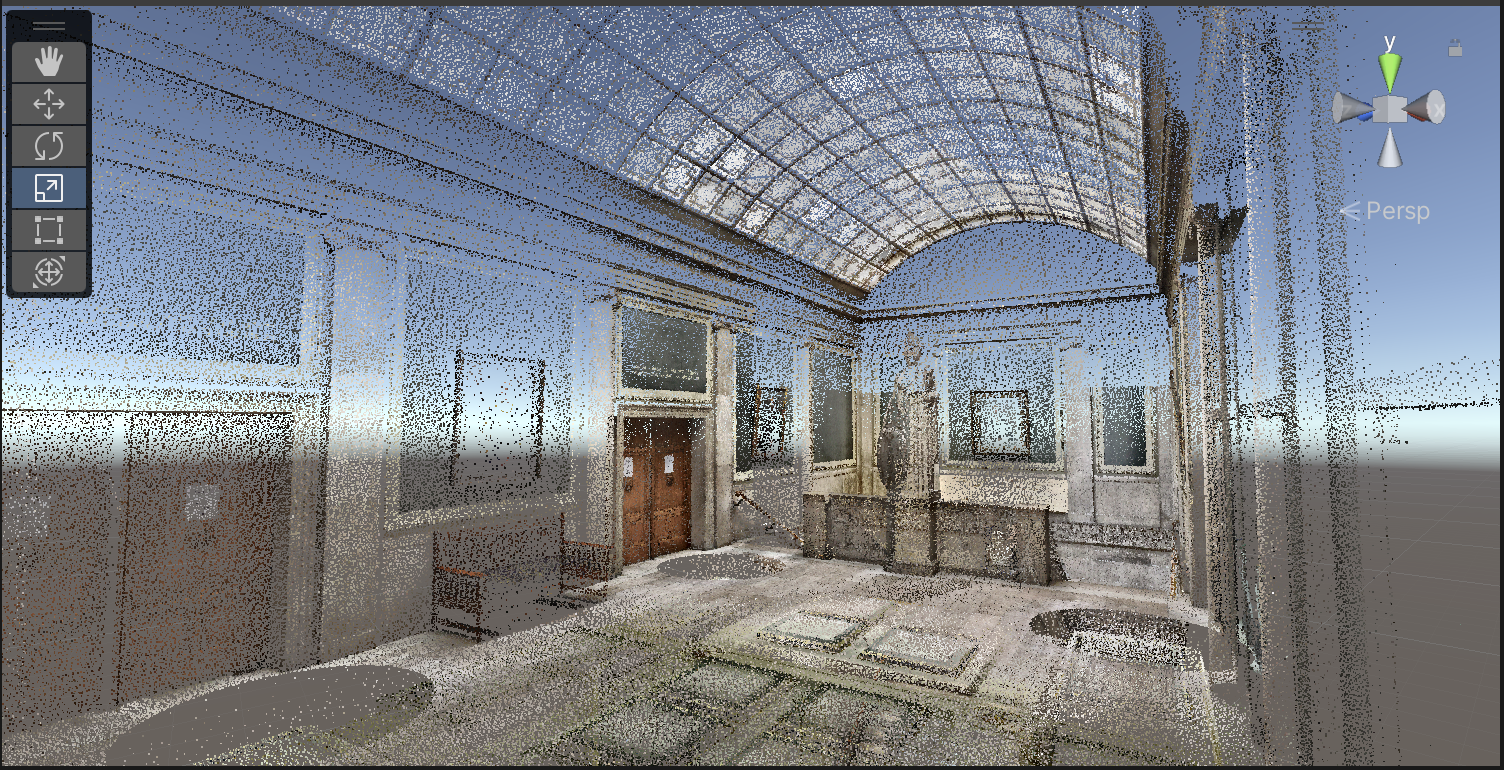}
    \caption{A decimated cloud loaded for the Great Hall cloud during octree construction}
\end{figure}

\begin{figure}
    \centering
    \includegraphics[width=\columnwidth]{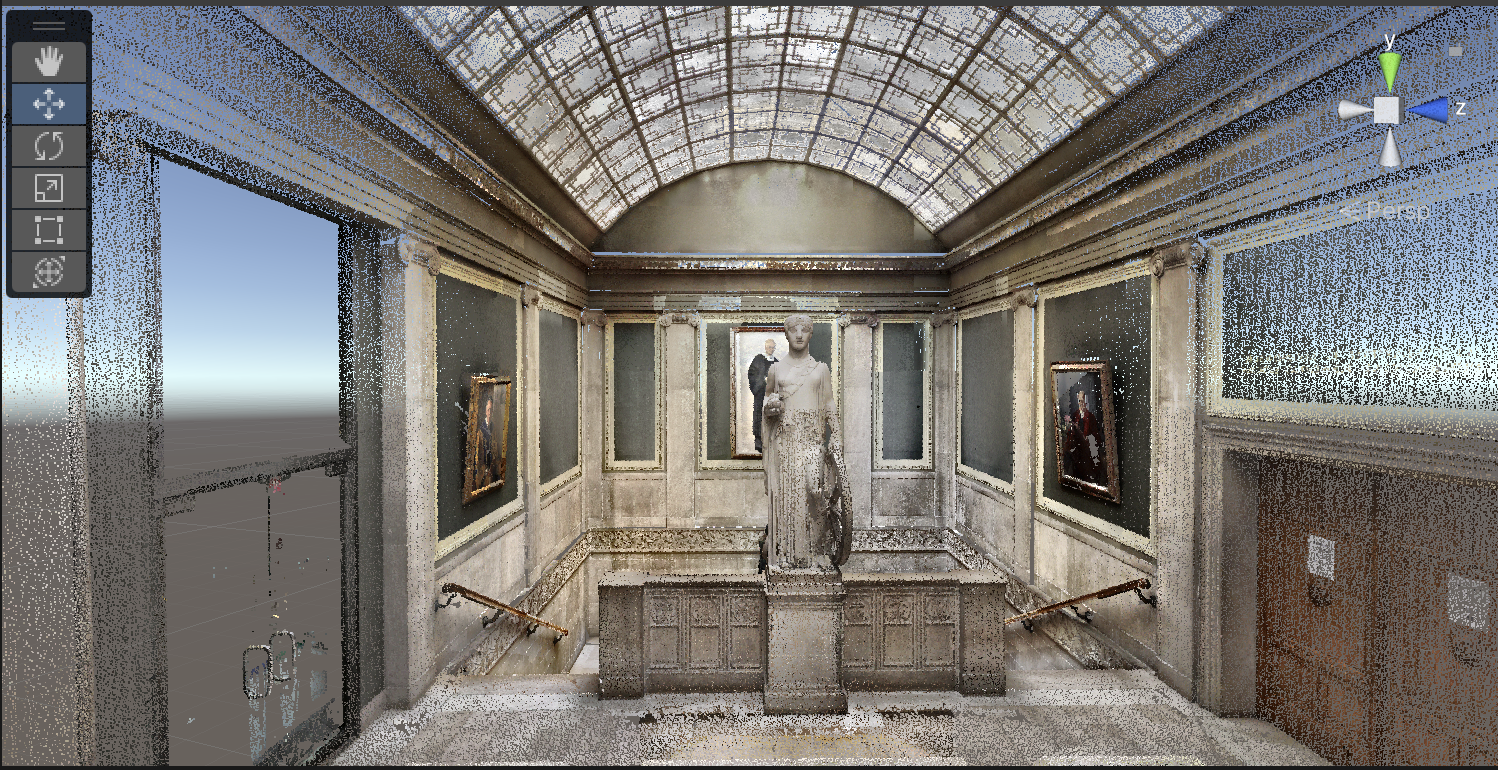}
    \caption{Full detail on the statue and surrounding furnishings after octree construction}
\end{figure}

\subsubsection{Integrable with standard mesh rendering pipelines}

Users today also want to integrate point cloud visualizations with other types of graphics. Allowing point clouds and meshes to coexist in virtual scenes gives users far more power to bring their visions to life. Even in scenes without meshes, connecting a point cloud renderer to a standard mesh rendering pipeline allows postprocessing, particle effects, and text elements to be used in conjunction with the cloud, creating more opportunities for educators, artists, and game developers to integrate point clouds into their work. FastPoints integrates fully with the Unity environment, allowing users to add meshes, particles, canvas elements, or any other graphical formats to their scenes.

\subsubsection{Interactive during preprocessing}

Another impediment to the ease of use of point cloud renderers is the lack of interactivity during the preprocessing step necessary for large point clouds. Many tools require an explicit preprocessing step, where users must install another program, run that on their unordered point cloud, and only then load the converted cloud into the renderer and visualize their cloud. While some other tools have an implicit preprocessing step, meaning the preprocessing is done by the same program that contains the renderer, these tools usually block the renderer while the data structure is built, so users are unable to interact with the application until this process is done. By providing interactivity during preprocessing, a renderer can enable users to integrate point clouds much more naturally and quickly with their workflows, incentivizing more experimentation and exploration. FastPoints uses a novel asynchronous program architecture to allow interactivity with a downsampled version of a user's point cloud while preprocessing is performed online.

\subsubsection{Support for popular point cloud formats}

Lastly, support for popular formats vastly increases the accessibility of a modern point cloud renderer. While LAS and LAZ files have seen a rise in popularity due to their specific design choices for storing point data, the PLY format is still very popular as well. If a renderer provides support for some and not others, users must seek out conversion tools that may require programming experience or tinkering to work properly. FastPoints supports PLY, LAS, and LAZ files natively.

\begin{table*}[b!]
    \centering
    \begin{tabular}{|c||c|c|c|c|c|}
        \hline
            Method & 
            \begin{tabular}{@{}c@{}}Large point \\ cloud support\end{tabular} & \begin{tabular}{@{}c@{}} Unity \\ Integration\end{tabular} & \begin{tabular}{@{}c@{}}Interactive during \\ preprocessing\end{tabular} &
            \begin{tabular}{@{}c@{}}Non-programmatic \\ interface\end{tabular} &
            \begin{tabular}{@{}c@{}}Support for \\ LAS, LAZ, \\ and PLY files\end{tabular}\\
        \hline
        pcx & \XSolid & \Checkmark & N/A & \Checkmark & \XSolid \\
        \hline
        Potree & \Checkmark & \XSolid & \XSolid & \XSolid & \XSolid \\
        \hline
        BA\_PointCloud & \Checkmark & \Checkmark & \XSolid & \XSolid& \XSolid \\
        \hline
        FastPoints (our method)& \Checkmark & \Checkmark & \Checkmark & \Checkmark& \Checkmark \\
        \hline
    \end{tabular}
    \caption{Table \ref{tab:features}: Feature Comparison of Existing Point Cloud Renderers}
    \label{tab:features}
\end{table*}

\begin{figure}[t]
    \centering
    \includegraphics[width=.9\columnwidth]{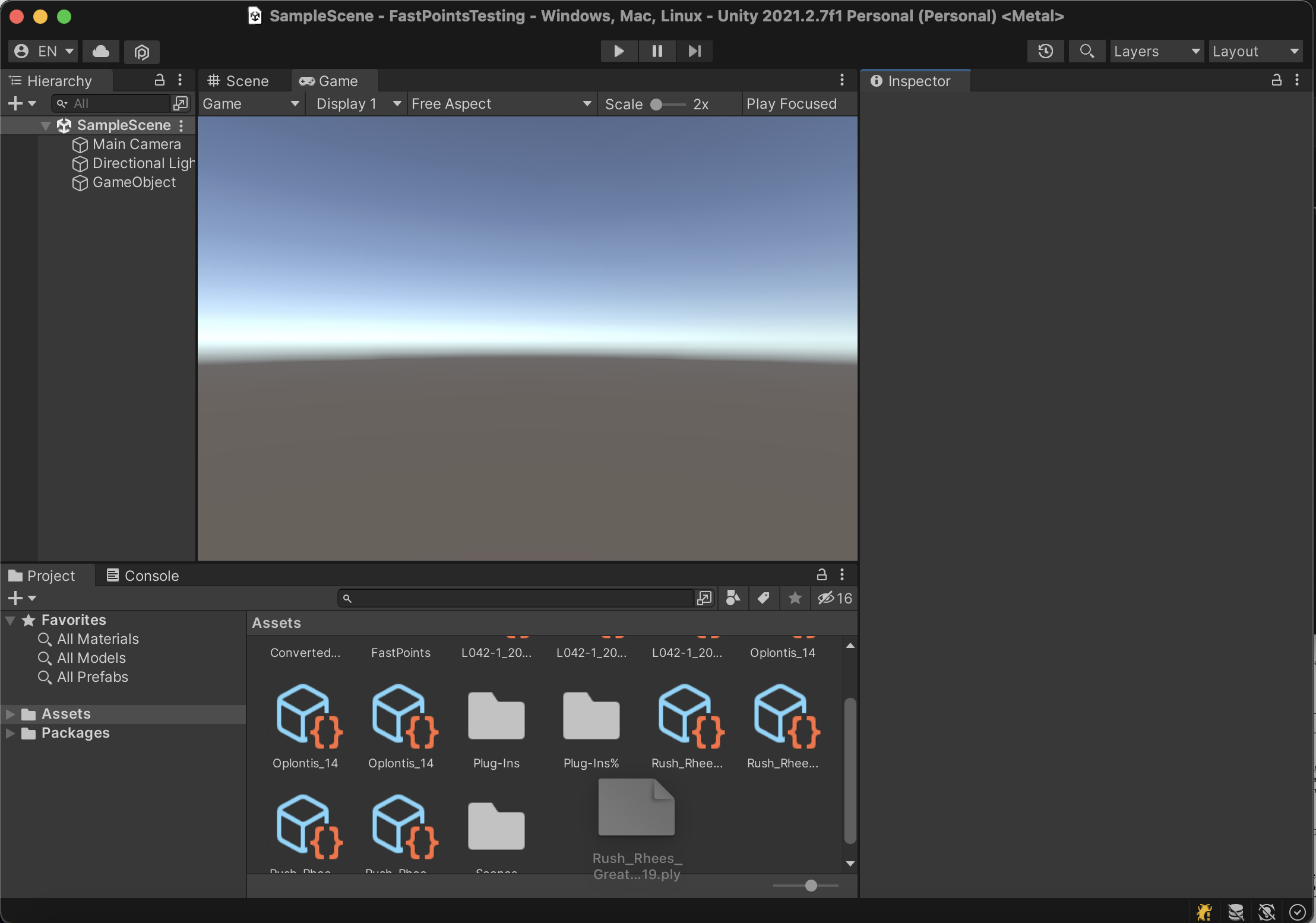}
    \caption{A point cloud being added to Unity via drag-and-drop. Other point cloud handles, stored as Unity ScriptableObjects, are visible in the Assets pane}
\end{figure}

\section{Prior Work}
Point cloud rendering has received an impressive amount of attention over the past decades. Of particular relevance to this project are previous papers discussing point cloud rendering pipelines, out-of-core data structures for large point clouds, and point cloud support in the Unity environment.

\subsection{Point Cloud Rendering}
As point cloud visualization has grown in popularity, researchers have identified a number of different avenues for improving rendering technology.

\subsubsection{Rendering pipelines}
A large volume of work has focused on optimal rendering pipelines for point clouds. In recent years, focus has shifted to leveraging GPGPU methods in these pipelines. Schütz et. al. propose a method to use compute shaders to pack per-pixel depth and color information into 64-bit integers and eliminate redundant fragment drawing on the final render pass \cite{schutz2021rendering}. Another GPGPU method was proposed by Günther et. al. in 2013 to perform a depth test with a busy-wait locking system and an additional early depth test to avoid waiting wherever possible \cite{gunther2013gpgpu}.

\subsubsection{GPU point loading}
Research has also been done on more efficient methods for loading point clouds to the GPU from disk or main memory. Schütz et. al. propose a method to progressively transfer points from main memory to the GPU, avoiding frozen frames during initial cloud loading, as well as a method to dynamically compute and render sub-clouds based on real-time camera positions through a combination of reprojection and repopulation of a separate rendered point buffer \cite{schutz2020progressive}.

\subsubsection{Visual fidelity}
A number of new methods have focused on increasing the visual quality of rendered point clouds. Many of these work to solve the surface or hole-filling problem \cite{surfaceproblem}, which is the inherent difficulty of displaying a convincing continuous surface from a point cloud without converting the cloud to a mesh or another explicit surface representation. Some solutions, such as the AutoSplats method \cite{autosplats}, display normal-aligned ellipsoids that can then be rasterized to create realistic surfaces - a process known as splatting. Other methods, such as Kashyap et. al. \cite{raytracing}, focus on raytracing points or splats, enabling global illumination and other realistic lighting effects. Still others work to improve the visual quality of particular rendering pipelines: continuous level-of-detail allows out-of-core tree structures to be rendered without jarring drops in resolution \cite{continuouslod}. Research has also been done on the effect of point cloud compression, a parallel problem to visual fidelity, on perceptual quality of renders \cite{coding}.

\subsection{Octree Construction}
There has also been a considerable amount of research into efficient point cloud storage formats. One of the most popular data structure generation tools developed in recent years is the Potree converter, which generates an out-of-core octree from standard LAS or LAZ file types. The most recent version of the converter was described in 2020 by Schütz et. al. \cite{schutz2020fast} It separates the octree construction process into an explicit chunking step - in which it sorts points hierarchically into a number of small out-of-core files - and a parallel indexing step, in which it loads multiple chunks simultaneously and constructs local octrees on multiple threads before finally stitching together a global tree.

\subsection{Unity Point Cloud Support}
Although Unity does not natively support most popular point cloud formats, the engine does have a built-in point renderer for meshes, which draws vertex data as points without incorporating triangles whatsoever. This allows a user to programatically populate a Unity Mesh object with point cloud data as vertices, but provides no GUI support or point cloud-specific rendering options.

One method that attempts to improve this support is the pcx plugin developed by Keijiro \cite{pcx}. This plugin provides drag-and-drop support for binary little-endian PLY files, but no other formats. It loads all points into main memory and renders them using Unity's built-in mesh point renderer or a custom point shader. While this plugin provides an intuitive user experience and support for one of the most popular and accessible point cloud formats, it does have a few drawbacks. For one, the size of point clouds it can render is limited both by a user's RAM and GPU. Large point clouds - like those often used in academic, medical, or professional settings - can be 50GB or larger in size, far larger than the main memory available on most platforms. Additionally, the shader implemented by this plugin loads the entire point cloud to GPU memory regardless of camera position, meaning point cloud sizes are also limited by GPU memory, and camera positions that lead to many points mapping to a single pixel slow rendering considerably.

Another method, developed by Simon Fraiss in his Bachelor's Thesis \cite{Fraiss2017RenderingLP}, provides support for the Potree 1.7 point cloud format \cite{schutz2016potree}. This method is able to offer a number of advantages due to the Potree format's hierarchical octree structure, including frustum culling, level-of-detail rendering, and support for clouds much larger than main or GPU memory with little to no latency even on mid-tier hardware. The most noticeable drawback of this method is the hefty preprocessing requirement: the plugin itself provides no drag-and-drop support and requires point clouds to be converted into the Potree 1.7 format via a separate preprocessing step - which can take hours for large clouds - before they can be loaded into Unity.

\section{Program Structure}
FastPoints introduces a novel program structure and a number of innovations to avoid some of the limitations of existing point cloud rendering solutions. We prioritize seamless user experience, immediate interactivity, support for popular point cloud formats, and tight integration with the Unity environment.

To support seamless user experience, we integrate our plugin with Unity's built-in asset pipeline to allow for drag-and-drop importing of PLY, LAS, or LAZ files, three of the most popular unstructured point cloud formats. Once a cloud has been imported, we provide immediate interactivity by uniformly downsampling the cloud and rendering the result. This gives users a rough approximation of their implemented cloud while further preprocessing is performed.

After the initial downsampling step, we generate an out-of-core octree using the Potree Converter 2.0 described by Schütz et. al. in 2020 \cite{schutz2020fast}. During this time, the Unity environment and the initial downsampled cloud remain fully interactive, allowing users to position their point clouds in larger scenes or continue to work on other aspects of their project.

Once the octree is fully generated, we discard the decimated point cloud and traverse the octree, dynamically loading points based on the camera position each frame.

\begin{figure}[h!]
    \centering
    \includegraphics[width=.9\columnwidth]{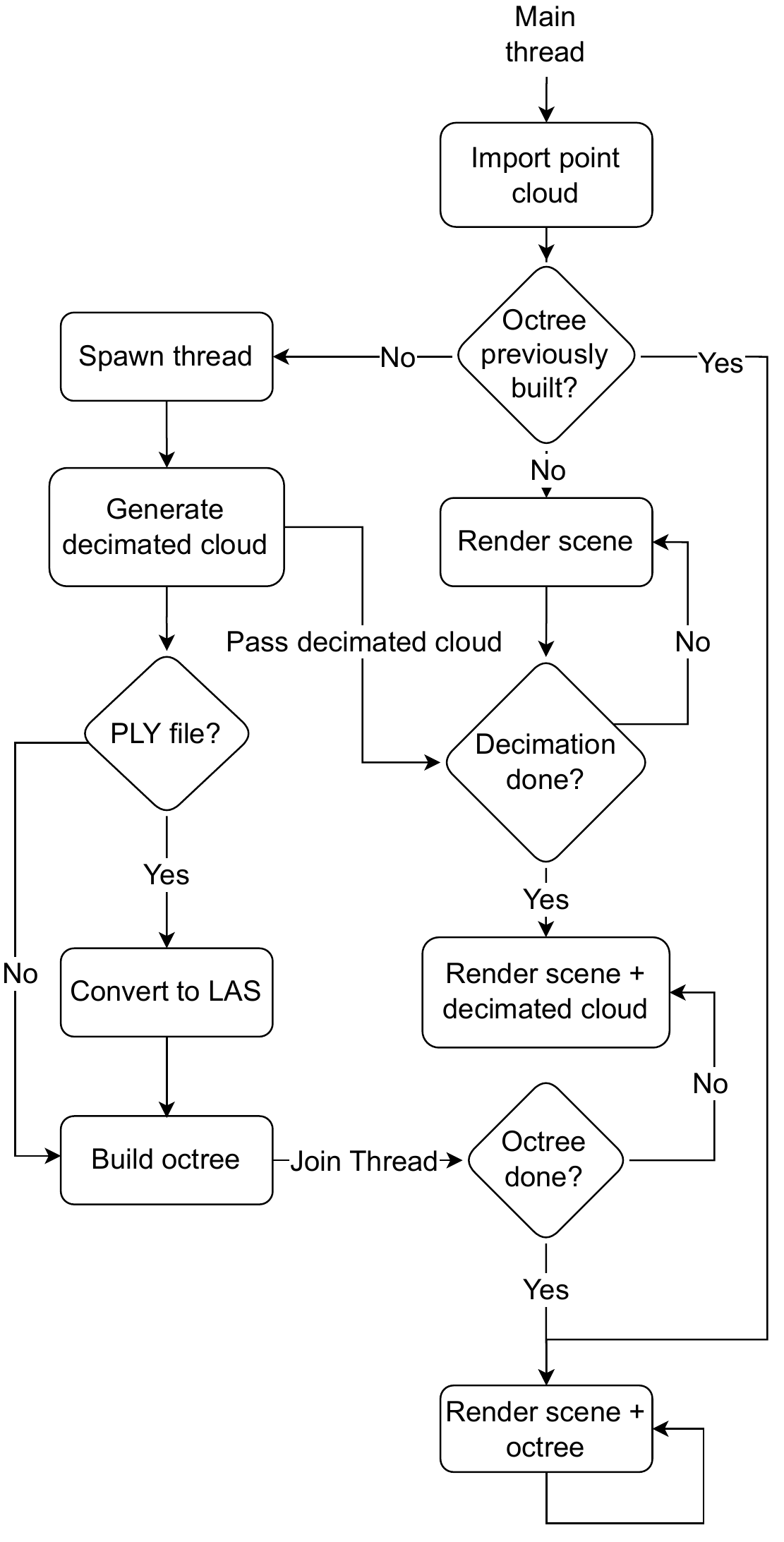}
    \caption{Our overall program structure. Point clouds are imported; then, if no octree has been constructed previously, the cloud is decimated for immediate rendering on the main thread and a new thread spawns to construct the octree. Once the tree is done, the decimated cloud is discarded and the octree is rendered in its place.}
    \label{fig:pipeline}
\end{figure}

The program structure is illustrated in Figure \ref{fig:pipeline}.

\section{Implementation}

Our tool utilizes a C\# plugin for interfacing with Unity as well as a native C++ plugin for most hefty IO operations and interaction with C++ conversion tools.

\subsection{File Importing}

For file importing, we utilize Unity's asset pipeline to support drag-and-drop importing of unordered point cloud PLY, LAS, or LAZ files. When a file is added, the plugin initializes a handle object for IO operations that can be passed to the C++ side of the plugin, which then starts conversion and octree construction operations.

\subsection{Immediate Low-Resolution Rendering}

Once the handle asset has been initialized, the program begins point cloud decimation. This uniform downsampling step requires iterating over the entire point cloud and thus has time requirements that are roughly linear with respect to point cloud size. To minimize the amount of time this step takes, we spawn multiple threads to subsample different chunks of the point cloud simultaneously and leverage the speed of modern SSDs. Since the time required for octree construction scales with cloud size as well, this subsampling step is consistently a small fraction of the total processing time, and usually finishes in a matter of seconds even for clouds with hundreds of millions of points.

One consideration with this step is the difference in performance between point cloud formats. While both PLY and LAS files can be iterated over and parsed quickly, LAZ files require a point-wise decoding to parse position and color attributes. This causes a significant time increase in decimation, but since this decoding is also necessary for octree construction from LAZ files, the total time for decimation remains a small proportion of the overall preprocessing time.

\subsection{File Conversion}

While the Potree Converter supports LAS and LAZ files out-of-the-box, it lacks support for PLY files. To address this, we include an additional conversion step for PLY files, using LAStools' las2las program to create temporary LAS files that can be used with the Potree Converter then cleaned up after the octree is constructed. This means processing times are longer for PLY files than LAS or LAZ files, but due to our program architecture, the decimated cloud is fully interactive for the duration of this additional conversion phase as well.

\subsection{Octree Construction}

\begin{figure*}[t]
    \centering
    \includegraphics[width=1.5\columnwidth]{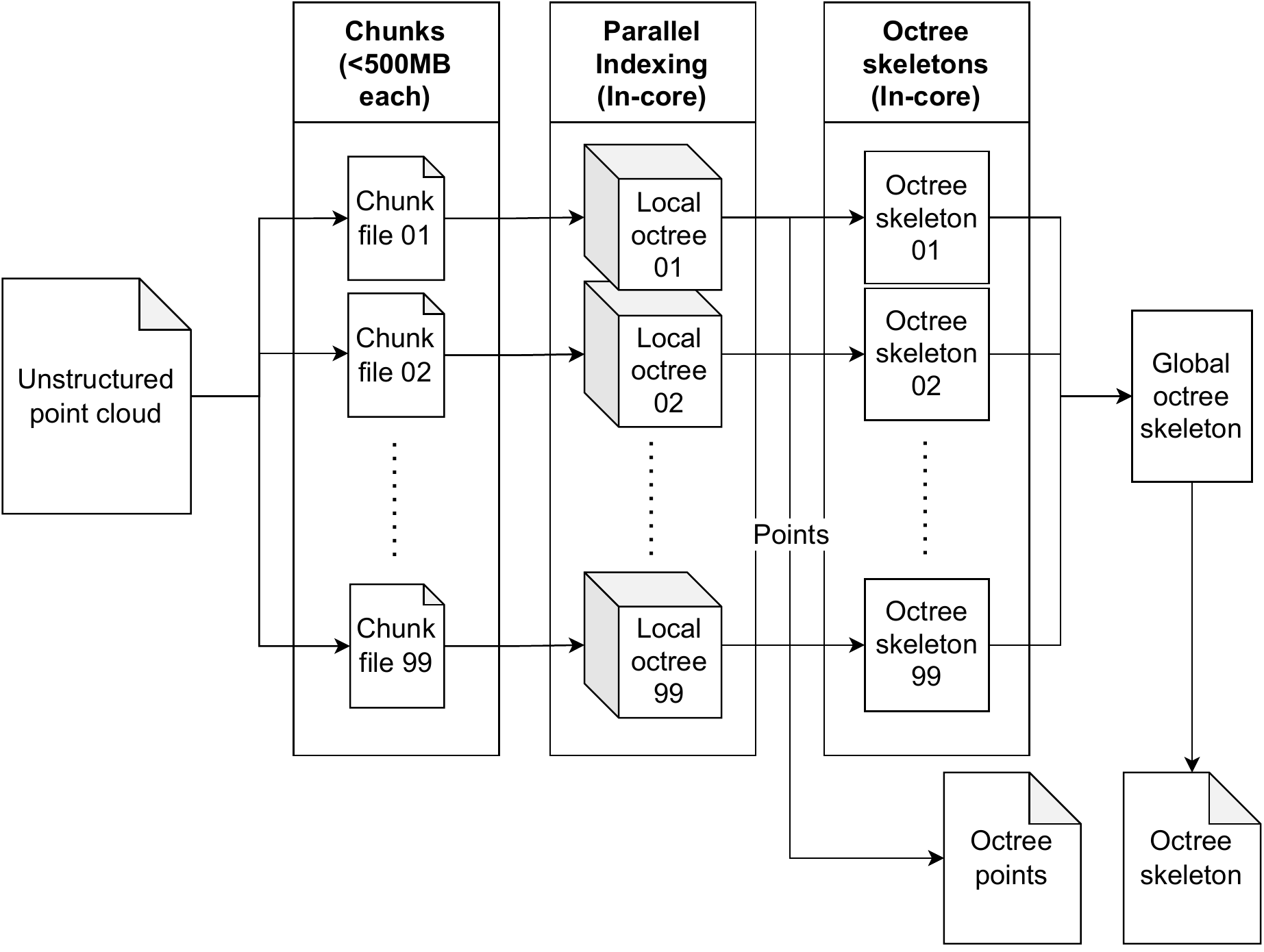}
    \caption{Octree construction pipeline, as described by Schütz et. al. The imported file is first sorted into a number of out-of-core chunk files, each representing a node in the global tree. The chunks are then indexed in parallel, creating local octrees, whose points are written to an additional file. Finally, the octrees are stitched together into a global tree, and the entire structure is written to an additional file.}
    \label{fig:octree}
\end{figure*}

Once the initial decimated cloud and conversion phase are done, an octree is constructed by the Potree Converter described in Schütz et. al. 2020 \cite{schutz2020fast}. First, the point cloud is sorted into a number of out-of-core chunks; then, the chunks are read back and converted into local octrees in parallel; finally, the local octrees are stitched together into a global octree that is written to a single file. This process is shown in Figure \ref{fig:octree}.

After construction is complete, the resulting files are linked to the created Unity asset, allowing the entire asset to be transferred between Unity projects seamlessly without loss of access to the original point cloud.

\subsection{Octree Rendering}

Once the entire octree is written to disk, the original decimated cloud is cleared from memory, and the octree skeleton is used to render points from the generated structure using the current camera position. The full rendering pipeline is shown in Figure \ref{fig:pipeline}.

To avoid slowing the main thread during octree traversal, we create a separate traversal thread that walks the tree and checks each node's visibility. If a node should be rendered, the traverser checks if its points are loaded from disk. If they are, the node is added to a queue of nodes to be rendered; if not, the node is marked and its points are asynchronously loaded into a Unity ComputeBuffer object allocated directly on the GPU. Conversely, if a node has points loaded but is not visible in the frame, its points are discarded to free GPU memory. Once the tree is fully traversed, the newly populated node queue is sent to the main thread.

\begin{figure*}[tbp]
    \centering
    \includegraphics[width=1.4\columnwidth]{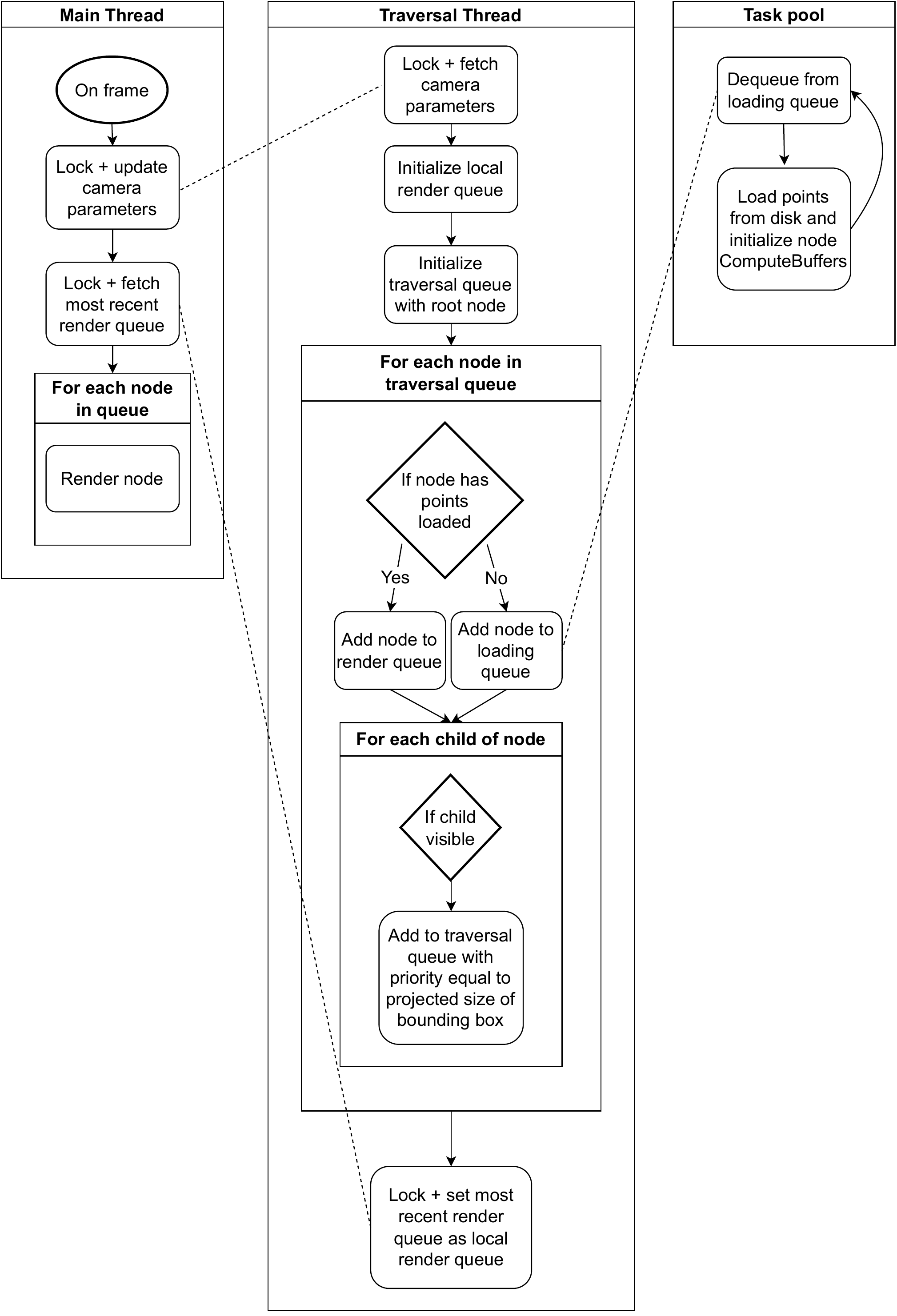}
    \caption{Overview of our CPU-side octree rendering algorithm. A spawned thread receives camera position and frustum data from the main thread, then traverses the octree, loading points into all nodes that are in-frame. The main thread, meanwhile, reads loaded points from the octree each frame and writes them to a GPU buffer.}
    \label{fig:render}
\end{figure*}

The main thread executes its work in the Unity per-frame Update and OnRenderObject methods. Each frame, the main thread sends frustum and positional camera information to the traverser thread, fetches the most recently sent node queue, and iterates through the queue. For each node, it performs a draw call directly with Unity's DrawProceduralNow method, avoiding the overhead of creating GameObjects for each node and allowing more direct access to Unity's rendering pipeline. The rendering pipeline is illustrated in Figure \ref{fig:render}.

\section{Evaluation}

Of our stated objectives - seamless user experience, immediate interactivity, and support for popular point cloud formats - only time to interactivity is directly quantifiable. To evaluate this, we measure the time taken by point cloud decimation on three clouds of different sizes.

We tested decimation on three point clouds: a 51 million-point NEON cloud \cite{DP1.30003.001/provisional}, a 222 million-point cloud of the Great Hall in University of Rochester's Rush Rhees Library collected by Michael Jarvis \cite{greathallmodel}, and the Oplontis 14 cloud of 989 million points also from Michael Jarvis's collection \cite{oplontismodel}. All tests were run on a 14" Macbook Pro 2021 with an M1 Pro CPU and 16GB RAM. Each cloud was decimated to 1,000,000 points in the decimation phase. The results can be seen in Table \ref{tab:decimationtimes}.

\begin{table*}[b]
\centering
\begin{tabular}{ |c|c||c|c|c| }
 \hline
 Cloud & Points & Decimation Time (PLY) & Decimation Time (LAS) & Decimation Time (LAZ)\\ 
 \hline
 NEON & 51,161,407 & 1469ms & 2670ms & 3186ms \\ 
 \hline
 Great Hall & 222,708,159 & 8611ms & 7068ms & 17.81s \\ 
 \hline
 Oplontis & 989,305,017 & 47.74s & 18.58s & 96.03s \\
 \hline
\end{tabular}
\caption{Table \ref{tab:decimationtimes}: Decimation times for three clouds in PLY, LAS, and LAZ formats}
\label{tab:decimationtimes}
\end{table*}

As expected, the decimation time requirements remained low in all three cases. While the measured times do not perfectly fit a linear time complexity, the requirements in each case were far less than the preprocessing requirements for generating a full out-of-core data structure. The full time requirements for each cloud in PLY format can be seen in Table \ref{tab:construction}.

\begin{table*}[b]
\begin{center}
\begin{tabular}{ |c|c||c|c|c| } 
 \hline
 Cloud & Points & Decimation Time & Conversion Time & Octree Construction Time \\ 
 \hline
 NEON & 51,161,407 & 1469ms & 7888ms & 29.91s \\ 
 \hline
 Great Hall & 222,708,159 & 8611ms & 1m03s & 2m54s \\ 
 \hline
 Oplontis & 989,305,017 & 47.74s & 4m50s & 10m40s \\
 \hline
\end{tabular}
\end{center}
\caption{Table \ref{tab:construction}: Processing times for three clouds in PLY format. In all cases, the initial decimation procedure is significantly faster than the processing as a whole}
\label{tab:construction}
\end{table*}

In addition to the increased time requirements for LAZ files, on large point clouds we also observe significantly higher decimation time requirements for PLY files than LAS files. This is due to the initial load requirements of PLY files being significantly below those of LAS files, but the per-point reading time after the initial load being considerably faster for the latter.

\begin{figure}[t]
    \centering
    \includegraphics[width=1\columnwidth]{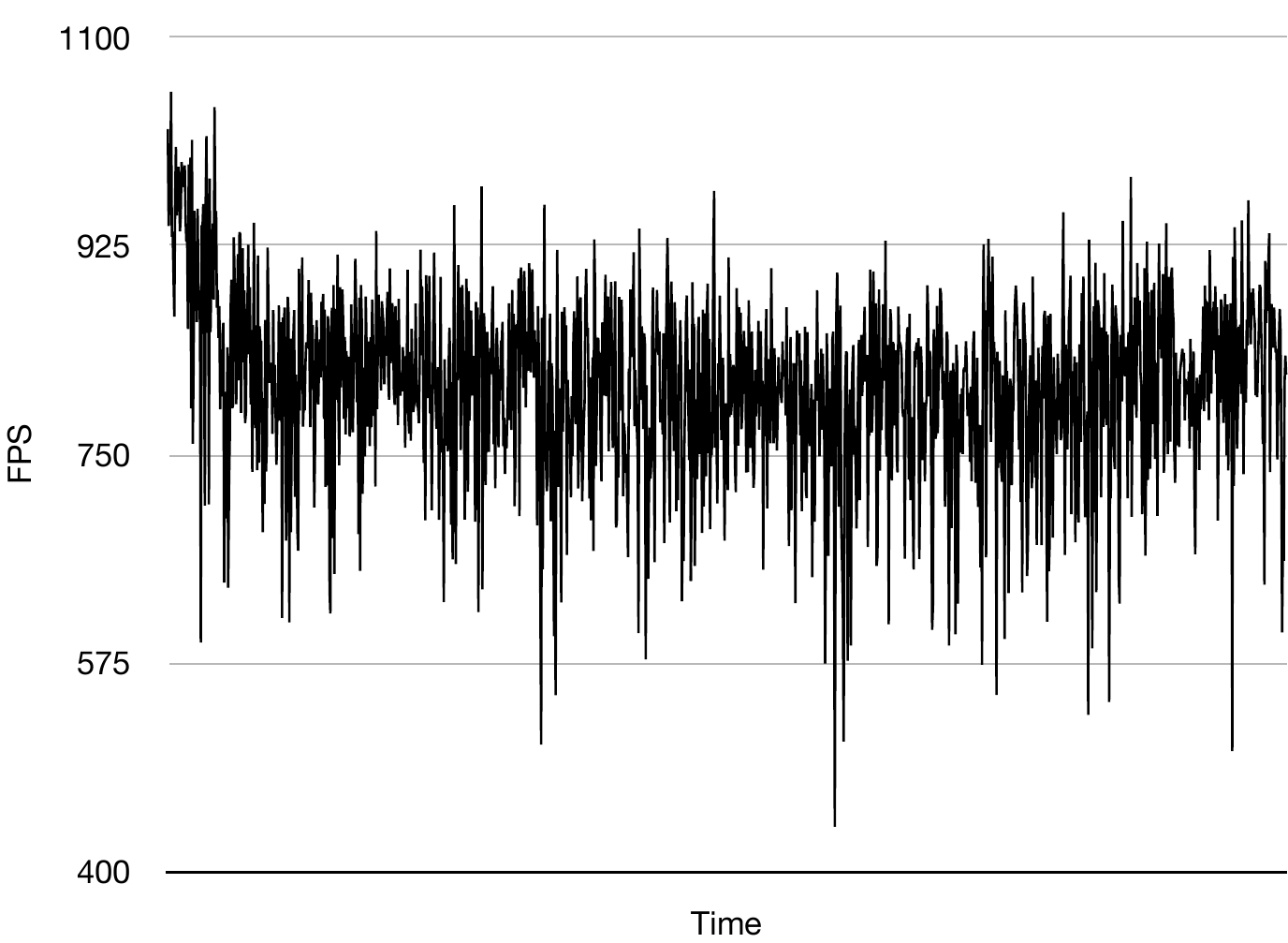}
    \caption{FPS during conversion and octree construction. The initial drop from near 1100 to closer to 800 occurs when the decimated cloud is first loaded}
    \label{fig:fps}
\end{figure}

We can also evaluate the interactivity of our solution by tracking frames per second (FPS) during conversion and octree construction. With a single camera orbiting the decimated Rush Rhees Great Hall cloud - consisting of 1 million points in its decimated state - we find on the same 14" Macbook Pro that the FPS never drops below four hundred, leaving plenty of rendering power for whatever other objects a user would want to add to their scene during this time. The FPS over time can be seen in Figure \ref{fig:fps}. 

\section{Future Work}

One advantage of the choice of Unity as a development platform is that it offers a robust suite of tools for extending FastPoints and improving the plugin in a variety of ways.

One intended expansion for FastPoints is the integration of the GPGPU rendering pipeline described by Günther et. al. \cite{gunther2013gpgpu} This would allow more points to be rendered per frame without overloading the GPU, meaning an octree with less granularity could see the same performance as more granular ones currently in-use in the application. These shallower trees would have lower construction times and would decrease the time needed until the full point cloud is interactive.

Due to Unity's powerful compute shader support, this could be implemented with the current rendering pipeline by replacing each node's draw call with a compute shader call that renders the node's points to a render texture, performing early z-culling and busy-waiting as described in the original paper. The final render texture would then be combined with the currently-rendering camera's color and depth textures to display the point cloud in context and with proper z-culling with the rest of the scene.

This shallow tree structure could be enhanced by the inclusion of progressive rendering as described by Schütz et. al. \cite{schutz2020progressive} This would allow significantly more points to be sent to the GPU without wasting as much rendering power, even when some points loaded to the GPU get clipped or occluded. The reproject-fill-prepare loop described in the paper would allow points to be reloaded from disk only as needed while the majority remain in GPU memory between frames.

Other planned features include adaptive point sizing, specifically for the initial decimated cloud, as well as point blending and eye dome lighting.

Additionally, while the primary focus of this tool thus far has been user experience and high-level design, we hope to extend the tool with a more performant low-level interface to enhance the speed and flexibility of the project even further. Specifically, the tool currently interfaces with LAStools on the native side to handle cloud decimation and PLY conversion. While this performs extremely well for decimating and processing LAS and LAZ files, results with PLY files could be improved. For instance, the PLY-to-LAS conversion step could be avoided completely with an implementation of the Potree Converter method that supports PLY files. Additionally, LAStools's PLY reader is designed to parse LAS attributes such as bounds that are unnecessary for the decimation step of our program; a custom implementation of a PLY reader could significantly reduce the time needed for subsampling PLY files.

More research could also be done in the sampling method for the initial decimated cloud. While uniform sampling was chosen for this application to maximize subsampling speed, one could imagine point clouds that are ordered such that uniform sampling results in unevenly distributed points in the decimated cloud. While none of the clouds we tested had this issue, if this becomes an issue in the future, it could be handled by a spatially-constrained random sampler.

\section{Conclusion}

In this paper, we present FastPoints, a new point cloud rendering plugin for the Unity development platform. This plugin offers constant interactivity, seamless user experience, support for massive, unordered clouds in popular formats, and tight integration with the Unity development platform. To achieve this, we introduce a novel architecture thxat uses an initial subsampling step to provide an immediately interactive estimation of the point cloud's geometry as well as integrated conversion and octree construction processes that proceed without blocking the Unity editor.

%\section{Acknowledgments} 
%I wish to thank Professor Zhu and Professor Jarvis for guiding me through this process and providing me with the resources I needed throughout development. I would also like to thank the University of Rochester Computer Science department and Joshua Romphf for introducing me to point cloud rendering in Unity.


%%%%%%%%%%%%%%%%%%%%%%%%%%%%%%%%%%
% Bibliography
%%%%%%%%%%%%%%%%%%%%%%%%%%%%%%%%%%

\bibliographystyle{abbrv}
\bibliography{template}
% \small
% \begin{thebibliography}{9}
% \bibitem{bib1}John Doe, Recent Progress in Digital Halftoning II,
  % IS\&T, Springfield, VA, 1999, pg. 173.
% \bibitem{bib2}John Doe, Digital Imaging, J. Imaging. Sci. and
  % Technol., 42, 112 (1998).
% \bibitem{bib3}John Doe, An Inexpensive Micro-Goniophotometry You Can
  % Build, Proc. PICS, pg. 179. (1998).
% \end{thebibliography}


\begin{biography}
Elias Neuman-Donihue received his BS in computer science and his BA in history from the University of Rochester (2022). He is currently working as a software developer at August Schools. His areas of interest include user experience, computer graphics, and web development.

Michael Jarvis is an Associate Professor of History, Director of Digital Media Studies Program, and Director of Smiths Island Archaeology
Project at the University of Rochester. He works onEarly American, Atlantic, Maritime, Public and Digital history and historical archaeology.
He obtained his Ph.D. from College of William and Mary (1998).

Yuhao Zhu received his BS in Computer Science from Beihang University (2010) and his Ph.D. in Electrical and Computer Engineering from
The University of Texas at Austin (2017). He is an Assistant Professor of
Computer Science at University of Rochester. His work focuses on applications, algorithms, and systems for visual computing
\end{biography}

\end{document}